\documentclass[aps,showpacs,preprintnumbers,preprint]{revtex4}
\usepackage{graphicx}

\begin{document}

\title{Generalized dark gravity\footnote{Essay written for the Gravity Research Foundation
2012 Awards for Essays on Gravitation}}
\author{Tiberiu Harko}
\email{harko@hkucc.hku.hk}
\affiliation{Department of Physics and
Center for Theoretical and Computational Physics, The University
of Hong Kong, Pok Fu Lam Road, Hong Kong, P. R. China}
\author{Francisco S. N. Lobo}
\email{flobo@cii.fc.ul.pt}
\affiliation{Centro de Astronomia e Astrof\'{\i}sica da
Universidade de Lisboa, Campo Grande, Ed. C8 1749-016 Lisboa,
Portugal}

\begin{abstract}

The late-time cosmic acceleration may be due to infra-red
modifications of General Relativity. In particular, we consider a
maximal extension of the Hilbert-Einstein action and analyze
several interesting features of the theory. Generally, the motion
is non-geodesic and takes place in the presence of an extra force,
which is orthogonal to the four-velocity. These models could lead
to some major differences, as compared to the predictions of
General Relativity or other modified theories of gravity, in
several problems of current interest, such as cosmology,
gravitational collapse or the generation of gravitational waves.
The study of these phenomena may also provide some specific
signatures and effects, which could distinguish and discriminate
between the various gravitational models.

\end{abstract}


\date{\today}

\maketitle


Cosmology is said to be thriving in a golden age, where a central
theme is the perplexing fact that the Universe is undergoing an
accelerating expansion \cite{Perlmutter,Riess1,Riess2}. The
latter, one of the most important and challenging current problems
in cosmology, represents a new imbalance in the governing
gravitational equations. Historically, physics has addressed such
imbalances by either identifying sources that were previously
unaccounted for, or by altering the governing equations. The cause
of this acceleration still remains an open and tantalizing
question. The standard model of cosmology has favored the first
route to addressing the imbalance, namely, a missing
energy-momentum component. In particular, the dark energy models
are fundamental candidates responsible for the cosmic expansion
(see \cite{Copeland:2006wr} for a review).

One may also explore the alternative viewpoint, namely, through a
modified gravity approach. It is widely believed that string
theory is moving towards a viable quantum gravity theory. In this
context, one of the key predictions of string theory is the
existence of extra spatial dimensions. In the brane-world
scenario, motivated by recent developments in string theory, the
observed 3-dimensional universe is embedded in a
higher-dimensional spacetime \cite{Maartens1}. Most brane-world
models, including those of the Randall-Sundrum type
\cite{Randall1,Randall2}, produce ultra-violet modifications to
General Relativity, with extra-dimensional gravity dominating at
high energies. However it is also possible for extra-dimensional
gravity to dominate at low energies, leading to infra-red
modifications of General Relativity. New features emerge in the
brane scenario that may be more successful in providing a
covariant infra-red modification of General Relativity, where it
is possible for extra-dimensional gravity to dominate at low
energies. The Dvali-Gabadadze-Porrati (DGP) models \cite{DGP}
achieve this via a brane induced gravity effect. The
generalization of the DGP models to cosmology lead to
late-accelerating cosmologies \cite{Deffayet}, even in the absence
of a dark energy field \cite{Maartens3}. This exciting feature of
``self acceleration'' may help towards a new resolution to the
dark energy problem, although this model deserves further
investigation as a viable cosmological model \cite{Lue}. While the
DGP braneworld offers an alternative explanation to the standard
cosmological model, for the expansion history of the universe, it
offers a paradigm for nature fundamentally distinct from dark
energy models of cosmic acceleration, even those that perfectly
mimic the same expansion history. It is also fundamental to
understand how one may differentiate this modified theory of
gravity from dark energy models. The DGP braneworld theory also
alters the gravitational interaction itself, yielding unexpected
phenomenological extensions beyond the expansion history. Tests
from the solar system, large scale structure, lensing all offer a
window into understanding the perplexing nature of the cosmic
acceleration and, perhaps, of gravity itself \cite{Lue}. The
structure formation \cite{Maartens4} and the inclusion of
inflation are also important requirements of DGP gravity, if it is
to be a realistic alternative to the standard cosmological model.

In the context of modified theories of gravity, infra-red
modifications to General Relativity have been extensively
explored, where the consistency of various candidate models, have
been analyzed (see \cite{Nojiri:2010wj} for a recent review). The
Einstein field equation of General Relativity was first derived
from an action principle by Hilbert, by adopting a linear function
of the scalar curvature, $R$, in the gravitational Lagrangian
density. However, there are no a priori reasons to restrict the
gravitational Lagrangian to this form, and indeed several
generalizations of the Einstein-Hilbert Lagrangian have been
proposed, including ``quadratic Lagrangians'', involving second
order curvature invariants such as $R^{2}$, $R_{\mu \nu }R^{\mu
\nu }$, $R_{\alpha \beta \mu \nu }R^{\alpha \beta \mu \nu }$,
$\varepsilon ^{\alpha \beta \mu \nu }R_{\alpha \beta \gamma \delta
}R_{\mu \nu }^{\gamma \delta }$, $C_{\alpha \beta \mu \nu
}C^{\alpha \beta \mu \nu}$, etc \cite{Lobo:2008sg}. The
physical motivations for these modifications of gravity were
related to the possibility of a more realistic representation of
the gravitational fields near curvature singularities and to
create some first order approximation for the quantum theory of
gravitational fields. In this context, a more general modification
of the Einstein-Hilbert gravitational Lagrangian density involving
an arbitrary function of the scalar invariant, $f(R)$, has been
extensively explored in the literature. Recently, a renaissance of
$f(R)$ modified theories of gravity has been verified in an
attempt to explain the late-time accelerated expansion of the
Universe \cite{Carroll:2003wy}.

In particular, a maximal extension of the Hilbert-Einstein action,
has recently been explored \cite{Harko:2010mv}, with the action
given by
\begin{equation}
S=\int f\left(R,L_m\right) \sqrt{-g}\;d^{4}x~,
\end{equation}
where $f\left(R,L_m\right)$ is an arbitrary function of the Ricci
scalar $R$, and of the Lagrangian density corresponding to matter,
$L_{m}$. The energy-momentum tensor of matter is defined as
\begin{equation}
T_{\mu \nu }=-\frac{2}{\sqrt{-g}}\frac{\delta \left(
\sqrt{-g}L_{m}\right)}{ \delta g^{\mu \nu }}\,.
\end{equation}
Thus, the gravitational field equation of $f\left( R,L_{m}\right)
$ gravity model is given by
\begin{eqnarray}\label{field}
&&f_{R}\left( R,L_{m}\right) R_{\mu \nu }+\left( g_{\mu \nu
}\nabla _{\mu }\nabla^{\mu } -\nabla
_{\mu }\nabla _{\nu }\right) f_{R}\left( R,L_{m}\right) \nonumber\\
&&-\frac{1}{2}\left[ f\left( R,L_{m}\right) -f_{L_{m}}\left(
R,L_{m}\right)L_{m}\right] g_{\mu \nu }=\frac{1}{2}
f_{L_{m}}\left( R,L_{m}\right) T_{\mu \nu }\,.
\end{eqnarray}
For the Hilbert-Einstein Lagrangian, $f( R,L_{m})
=R/2\kappa^2+L_{m}$, we recover the standard Einstein field
equations of General Relativity, i.e., $R_{\mu \nu }-(1/2)g_{\mu
\nu }R=\kappa^2 T_{\mu \nu }$. For $f\left( R,L_{m}\right)
=f_{1}(R)+f_{2}(R)G\left( L_{m}\right) $, where $f_{1}$, $f_{2}$
and $G$ are arbitrary functions of the Ricci scalar and of the
matter Lagrangian density, respectively, we reobtain the field
equations of modified gravity with an arbitrary curvature-matter
coupling, considered in
\cite{Harko:2008qz,Bertolami:2007gv,Harko:2010hw}.

These models possess extremely interesting properties. First, the
covariant divergence of the energy-momentum tensor is non-zero,
and is given by
\begin{eqnarray}
\nabla ^{\mu }T_{\mu \nu }=2\nabla
^{\mu }\ln \left[ f_{L_m}\left(R,L_m\right) \right] \frac{\partial L_{m}}{%
\partial g^{\mu \nu }}\,.  \label{noncons}
\end{eqnarray}
The requirement of the conservation of the energy-momentum tensor
of matter, $\nabla ^{\mu }T_{\mu \nu }=0$, yields an effective
functional relation between the matter Lagrangian density and the
function $f_{L_m}\left(R,L_m\right)$, given by $\nabla ^{\mu }\ln
\left[ f_{L_m}\left(R,L_m\right) \right] \partial L_{m}/
\partial g^{\mu \nu }=0$.

In second place, the motion of test particles is non-geodesic, and
takes place in the presence of an extra force, orthogonal to the
four-velocity. As a specific example, consider the case in which
matter, assumed to be a perfect thermodynamic fluid, obeys a
barotropic equation of state, with the thermodynamic pressure $p$
being a function of the rest mass density of the matter $\rho $
only, so that $p=p\left( \rho \right) $. In this case, the matter
Lagrangian density, which in the general case could be a function
of both density and pressure, $L_{m}=L_{m}\left( \rho ,p\right) $,
or of only one of the thermodynamic parameters, becomes an
arbitrary function of the density of the matter $\rho $ only, so
that $L_{m}=L_{m}\left( \rho \right) $ (for more details, we refer
the reader to
\cite{Harko:2010mv,Bertolami:2008ab,Bertolami:2008zh}). Thus, the
equation of motion of a test fluid in $f\left(R,L_m\right)$
gravity is
\begin{equation}
\frac{d^{2}x^{\mu }}{ds^{2}}+\Gamma _{\nu \lambda }^{\mu }u^{\nu
}u^{\lambda }=f^{\mu },  \label{eqmot}
\end{equation}
where
\begin{equation}
f^{\mu }=-\nabla _{\nu }\ln \left[  f_{L_m}\left(R,L_m\right) \frac{%
dL_{m}\left( \rho \right) }{d\rho }\right] \left( u^{\mu }u^{\nu
}-g^{\mu \nu }\right) .
\end{equation}
The extra-force $f^{\mu }$  is perpendicular to the four-velocity,
$u^{\mu}$, i.e., $f^{\mu }u_{\mu }=0$.

The non-geodesic motion, due to the non-minimal couplings present
in the model, implies the violation of the equivalence principle,
which is highly constrained by solar system experimental tests
\cite{Faraoni,BPT06}. However, it has recently been argued, from
data of the Abell Cluster A586, that the interaction between dark
matter and dark energy implies the violation of the equivalence
principle \cite{BPL07}. Thus, it is possible to test these models
with non-minimal couplings in the context of the violation of the
equivalence principle. It is also important to emphasize that the
violation of the equivalence principle is also found as a
low-energy feature of some compactified versions of
higher-dimensional theories.

An interesting application of the latter $f(R,L_\mathrm{m})$
gravity is the $f(R,T)$ model proposed in \cite{Harko:2011kv},
where the action takes the following form
\begin{equation}
S=\frac{1}{16\pi}\int
f\left(R,T\right)\sqrt{-g}\;d^{4}x+\int{L_\mathrm{m}\sqrt{-g}\;d^{4}x}\,.
\end{equation}
$f\left(R,T\right)$ is an arbitrary function of the Ricci scalar,
$R$, and of the trace $T$ of the energy-momentum tensor of the
matter, $T_{\mu \nu}$. $L_\mathrm{m}$ is the matter Lagrangian
density. Note that the dependence from $T$ may be induced by
exotic imperfect fluids or quantum effects (conformal anomaly).

As a specific case of a $f(R,T)$ modified gravity model, consider
$f(R,T)=R+2f(T)$, where $f(T)$ is an arbitrary function of $T$. In
a cosmological setting, a simple model can be obtained by assuming
a dust universe ($p=0$, $T=\rho $), and by choosing the function
$f(T)$ so that $f(T)=\lambda T$, where $\lambda $ is a constant.
Considering a flat Robertson-Walker metric, $
ds^2=dt^2-a^2(t)\left(dx^2+dy^2+dz^2\right)$, the gravitational
field equations are given by
\begin{eqnarray}
3\frac{\dot{a}^2}{a^2}&=&\left(8\pi +3\lambda \right)\rho\, , \\
2\frac{\ddot{a}}{a}+\frac{\dot{a}^2}{a^2}&=&\lambda \rho\, ,
\end{eqnarray}
respectively. Thus, this $f(R,T)$ gravity model is equivalent to a
cosmological model with an effective cosmological constant
$\Lambda _\mathrm{eff}\propto H^2$, where $H=\dot{a}/a$ is the
Hubble function. It is also interesting to note that generally for
this choice of $f(R,T)$ the gravitational coupling becomes an
effective and time dependent coupling of the form
$G_\mathrm{eff}=G\pm 2f'(T)$ (see \cite{Harko:2011kv} for more
details). Thus the term $2f(T)$ in the gravitational action
modifies the gravitational interaction between matter and
curvature, replacing $G$ by a running gravitational coupling
parameter. The field equations reduce to a single equation for
$H$,
\begin{equation}
2\dot{H}+3\frac{8\pi +2\lambda }{8\pi +3\lambda }H^2=0\, ,
\end{equation}
with the general solution given by

\begin{equation}
H(t)=\frac{2\left(8\pi +3\lambda \right)}{3\left(8\pi +2\lambda
\right)}\frac{1}{t}\, .
\end{equation}
The scale factor evolves according to $a(t)=t^{\alpha }$, with
$\alpha ={2\left(8\pi +3\lambda \right)/3\left(8\pi +2\lambda
\right)}$.

In conclusion, the predictions of the maximal extensions of
General Relativity, namely the $f(R,L_m)$ gravity models could
lead to some major differences, as compared to the predictions of
standard General Relativity, or other generalized gravity models,
in several problems of current interest, such as cosmology,
gravitational collapse or the generation of gravitational waves.
The study of these phenomena may also provide some specific
signatures and effects, which could distinguish and discriminate
between the various gravitational models. In order to explore in
more detail the connections between the $f(R,L_m)$ gravity model
and the cosmological evolution, some explicit physical models are
necessary to be built.

\section*{Acknowledgments}

TH is supported by an RGC grant of the government of the Hong Kong
SAR. FSNL acknowledges financial support of the Funda\c{c}\~{a}o
para a Ci\^{e}ncia e Tecnologia through the grants
CERN/FP/123615/2011 and CERN/FP/123618/2011.



\end{document}